\documentclass{scrartcl}
\usepackage[utf8]{inputenc}
\usepackage{geometry}
\usepackage{graphicx}
\usepackage{amsmath}
\usepackage{amsthm}
\usepackage{ragged2e}
\usepackage[ruled,vlined]{algorithm2e}
\usepackage{float}
\usepackage{amssymb}
\usepackage{enumitem}
\usepackage{bbm}
\usepackage[english]{babel}
\usepackage{csquotes}
\usepackage{mathtools}
\usepackage{multicol}
\usepackage{appendix}
\newcommand\numberthis{\addtocounter{equation}{1}\tag{\theequation}}

\title{Portfolio Management and Return Prediction}
\subtitle{On Markowitz MPT, Constant Correlation Model, Single Index Model, Multi-factor Model}
\author{Qingyin Ge, Yunuo Ma, Rongyu Li, Yuezhi Liao, Tianle Zhu}
\date{May 2020}

\begin{document}
\maketitle
\begin{multicols}{2}

\section{Introduction}
With the well development in financial industry, market starts to catch people's eyes, not only by the diversified investing choices ranging from bonds and stocks, to futures and options, but also by the general "high-risk, high-reward" mindset prompting people to put money in financial market. What we always show concerns for, is nothing but two terms, risk, and return. People are interested in reducing risk at a given level of return since there is no way having both high return and low risk. Many researchers have been studying on this issue, and the most pioneering one is Harry Markowitz's \textit{Modern Portfolio Theory} developed in 1952, which is the cornerstone of investment portfolio management. Markowitz's MPT is one of the most widely-used structure in terms of portfolio construction, which aims at "maximum the return at the given risk". In contrast to that, fifty years later, E. Robert Fernholz's \textit{Stochastic Portfolio Theory}, as opposed to the normative assumption served as the basis of earlier modern portfolio theory, is consistent with the observable characteristics of actual portfolios and markets. \par 
In this paper, you will see first some basic theories of Markowitz's MPT and Fernholz's SPT. Next we step across to application side, trying to figure out under four basic models based on \textit{Markowitz Efficient Frontier}, including \textit{Markowitz Model}, \textit{Constant Correlation Model}, \textit{Single Index  Model}, and \textit{Multi-Factor Model}, what portfolios will be selected and how do these portfolios perform in real world. Here we also involve \textit{Universal Portfolio Algorithm} by Thomas M. Cover to select portfolios as comparison. In addition, each portfolio's \textit{Value at Risk}, \textit{Expected Shortfall} and corresponding \textit{Bootstrap confidence interval} for risk management will be evaluated. Finally, by utilizing factor analysis and time series model, we could predict future performance of our four models.

\section{Background Theory}
\subsection{Markowitz Modern Portfolio Theory}
\textit{Modern Portfolio Theory} assumes that investors are risk averse, where the risk is measured by the variance of asset price. It basically describes the "trade-off" between return and risk. Investors who want higher return must accept higher risk, but different people may have their own risk toleration, which lead to different investment strategies, forming the so-called \textit{Efficient Frontier}. Under the model, we can represent the expected portfolio return as
\begin{equation}
\mathop{\mathbb{E}}(R_p) = \sum_{i = 1}^{n}{w_i\mathop{\mathbb{E}}(R_i)} 
\end{equation}
where $w_i$ is the weight of asset i and $R_i$ is the corresponding asset return; the portfolio volatility as
\begin{equation}
\sigma_p^2 = \sum_{i = 1}^{n}{w_i^2\sigma_i^2} + \sum_{i}\sum_{j \neq i}{w_iw
_j\sigma_i\sigma_j\rho_{ij}}
\end{equation}
where $\sigma_i$ is the individual volatility of asset i and $\rho_{ij}$ is the correlation coefficient between returns on asset i and asset j. \par
If risk-free asset gets involved, we are stepping into \textit{Capital Asset Pricing Model}, which is so-called "CAPM". CAPM provides us a decent way to fairly price portfolios. We will use \textit{tangent portfolios} along with our four models for future investigation. \par

\subsection{Stochastic Portfolio Theory}
\textit{Stochastic Portfolio Theory} basically shows that  "the growth rate of a portfolio depends not only on the growth rates of the component stocks, but also on the \textit{excess growth rate}, which is determined by the stock's variances and covariances."(R. Fernholz and I. Karatzas, 2008). The stock capitalisations are modeled by \textit{Ito Process}, dynamically. Roughly speaking, n positive stock capitalisation processes $X_i$ can be modelled as follows
\begin{equation}
dX_i(t) = X_i(t)\left( r_i(t)dt + \sum_{\nu = 1}^d{\sigma_{i\nu}(t)dW_{\nu}(t)}\right)
\end{equation}
for $t \geq 0$ and $i = 1, ..., n$. Here $W_i$ are independent standard Brownian Motions and $X_i$ are capitalisations. Notice that this process is on logarithm scale since SPT uses geometric rate of return instead of arithmetic growth rate \cite{Fernholz}. It is also worth to mention that $r_i$ and $\sigma_i$ are \textit{$\mathop{\mathbb{F}}$-progressive} and satisfy sum of integral finite almost surely \cite{WS}. In addition, Fernholz and Shay (1982) were the first to observe that portfolio diversification and market volatility behave as drivers of a growth in such a frame. The growth of a well-diversified portfolio will dominate strictly the average of the individual assets growth rate. What really help is its application in machine learning framework, especially \textit{Functionally Generated Portfolios}. Consider a class of function $\mathbb{G} \in C^2(U, \mathbb{R}_+)$ with U an open set. \textit{Fernholz's Master Equation} is a pathwise decomposition of the relative performance of specific portfolios and that of market, which is free from stochastic integrals: 
\begin{equation}
log\left( \frac{X^\pi(T)}{X^\mu(T)}\right) = log\left( \frac{\mathbb{G}(\mu(T))}{\mathbb{G}(\mu(0))}\right) + \int_0^T g(t)dt
\end{equation}
where $g(\cdot)$ is called the drift process of the portfolio $\pi(\cdot)$. $\mathbb{G}$ is said to be the generated function of the functionally generated portfolio $\pi(\cdot)$. Furthermore, one of the most studied FGP is the \textit{diversity-weighted portfolios} (DWP) with parameter $p$ and some continuous function $f$ for long only, 
$$ \pi_i^{f}(t) := \frac{f(x_i(t))}{\sum_{j = 1}^n f(x_j(t))}, \quad i = 1, ..., n$$
As Y-L Kom Samo and A. Vervuurt (2016) mentioned, it is verified by real data that these portfolios have potential to outperform the market index, as well as their positive parameter counterparts \cite{Vervuurt}. We try to learn the investment strategy by learning the map $f  :  \mu \mapsto \mu^p, \; p \in [-1, 1]$, by evaluating the portfolios' sharpe ratio (SR) or the excess return (ER) relative to benchmark portfolio $\pi^*$(here is the equally weighted portfolio). We would like to maximize the objective functions
\begin{align*}
P_D(log(f)) =& SR(\pi) \\
=& \sqrt{252} \frac{\hat{E}(r(1), ..., r(T))}{\hat{S}(r(1), ..., r(T))}
\end{align*}
or
\begin{align*}
P_D(log(f)) =& ER(\pi^f | EWP) \\
=& \prod_{t = 1} ^ T (1 + \sum_{i = 1}^n r_i(t)\pi_i^f(t)) \\
-& \prod_{t = 1} ^ T (1 + \sum_{i = 1}^n r_i(t)\pi_i^*(t))
\end{align*}
where $\hat{E}$ is sample mean and $\hat{S}$ is sample sd. As a result, we will be able to catch $\underset{p}{\mathrm{argmax}} \;  P_D(log(f))$ and thus get the best investment strategy.

\section{Application}
\subsection{Data Overview}
We have chosen 8 stocks which belong to 6 different sectors from \textbf{Yahoo Finance}, including Amazon, Apple, Caterpillar, Delta, Google, JP Morgan, Tesla, and Mobil. Our data contain daily prices over the time period from Jan 1, 2011, to Dec 31, 2019, summing up to total $N = 2262$ observations.
\begin{figure}[H]
  \centering
  \includegraphics[width=8cm, height=2.5cm]{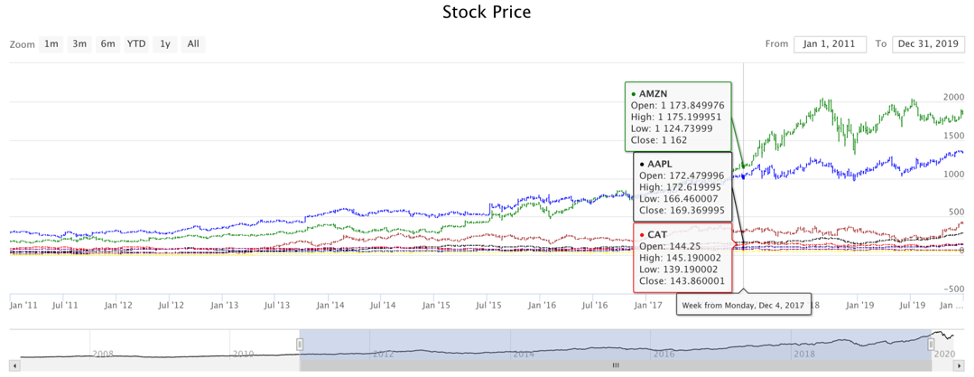}
  \caption{Daily Price of 8 Stocks from 1/1/2011 to 12/31/2019} 
\end{figure}
From QQ plots and histograms we find out that the returns are approximately "normally" distributed but with heavy tails, meaning that cases with unexpected high or low values are significantly more extreme than what would be expected from a normal distribution. Thus we believe the stock return follows more or less a t-distribution.

\begin{figure}[H]
  \centering
  \includegraphics[width=8cm, height=2.2cm]{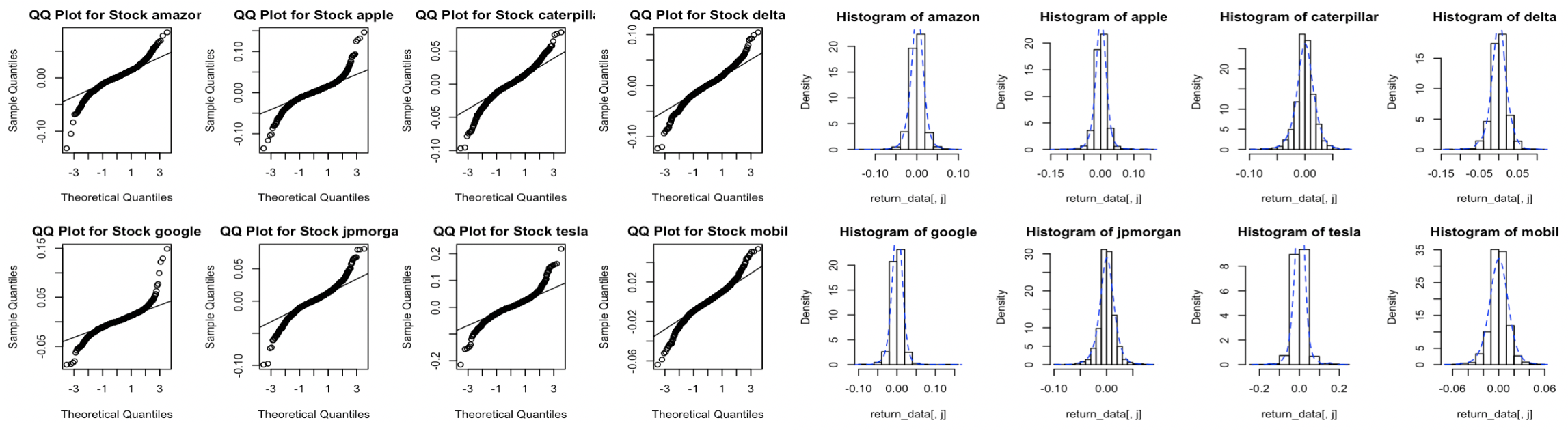}
  \caption{Daily Price of 8 Stocks from 1/1/2011 to 12/31/2019} 
\end{figure}

The correlation plot suggests that each pair of stocks has around 0.3 correlation between each other, which lead to careful consideration about the portfolio volatility structure.

\subsection{Risk Management}

\subsubsection{Universal Portfolio Algorithm}
Let's begin with \textit{Universal Portfolio Algorithm}. "Universal" is in the sense of no statistical assumptions underlying the market behavior, therefore the constructed portfolio is robust to real world market movements. There are mainly four of them:
\begin{itemize}
  \item Constant Rebalanced Portfolio (CRP), which uses 1/n as the portfolio weight and rebalanced it at the beginning of each trading period
  \item Cover Universal Portfolio (CUP), which calculates weights as
  \begin{equation}
      \hat{w_k} = \frac{\int wS_{k-1}(w)\pi(dw)}  {\int S_{k-1}(w) \pi(dw)}
  \end{equation} 
  where $S$ is the total wealth at current position
  \item Weighted Average of Best CRP, which calculates current portfolio weights as a weighted average of the historical best CRP until now
  \item Successively Best CRP, which is a momentum-based strategy using past weight for best CRP for next period. As ${\eta} \rightarrow \infty$
  \begin{equation}
      \hat{w_k}^{\eta} = \frac{\int w[S_{k-1}(w)]^{\eta}\pi(dw)}  {\int [S_{k-1}(w)]^{\eta} \pi(dw)}
  \end{equation}
  is the weight for SCRP
  
\end{itemize}
Based on our data, if we invest the initial wealth \$1 on 1/3/2011, the best asset Tesla will bring us \$15.57 on 12/31/2019, while the worst asset Caterpillar will only bring us \$1.25.

\begin{figure}[H]
  \centering
  \includegraphics[width=8cm, height=4.8cm]{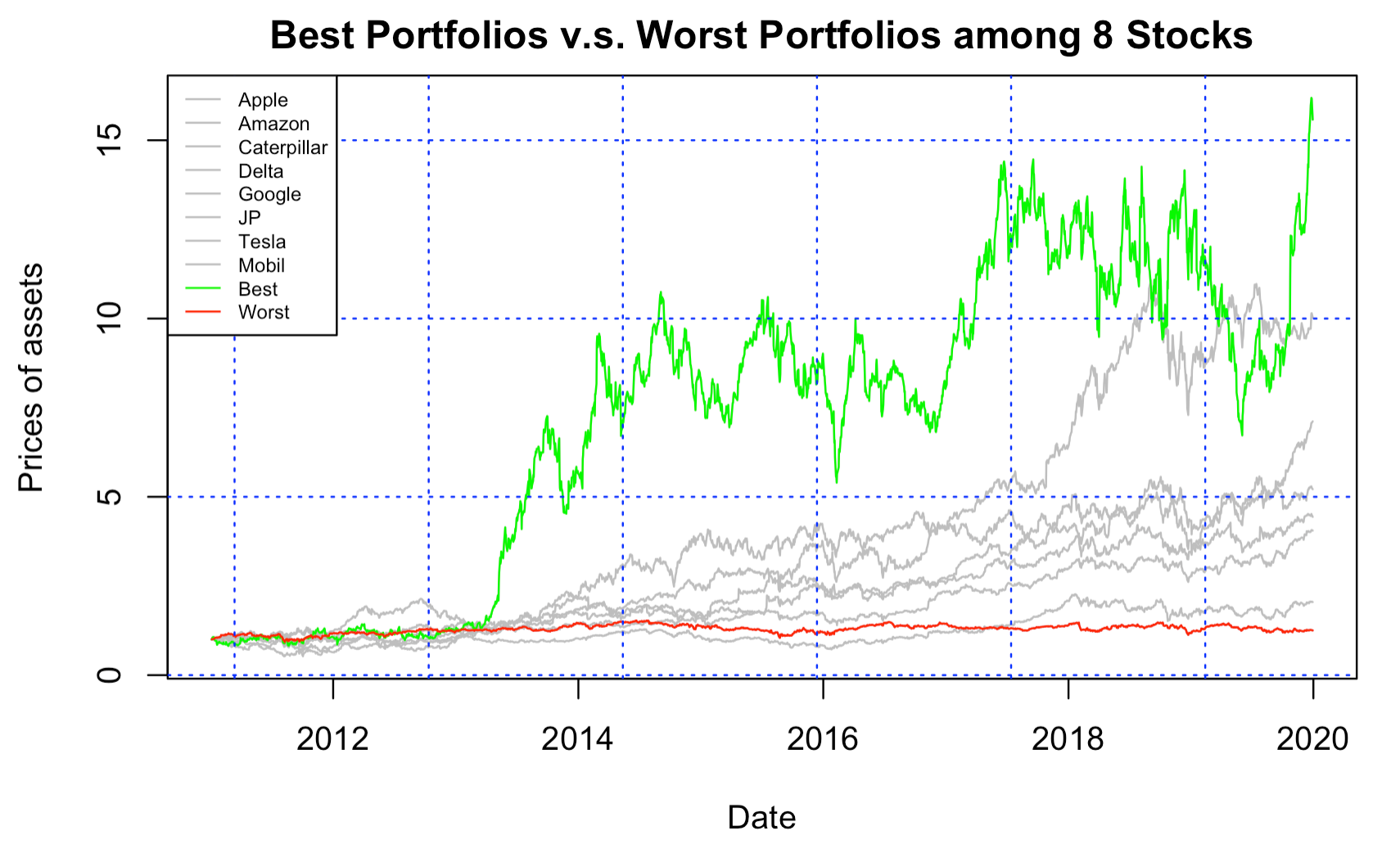}
  \caption{Best Asset and Worst Asset by CRP} 
\end{figure}

Via this algorithm, we could build four rebalanced portfolios. The result indicates that CUP, SCRP, CRP provide us better result, which make us end up with \$6 or so; while the weighted average CRP only give us \$4.5. 
\begin{figure}[H]
  \centering
  \includegraphics[width=8cm, height=3.5cm]{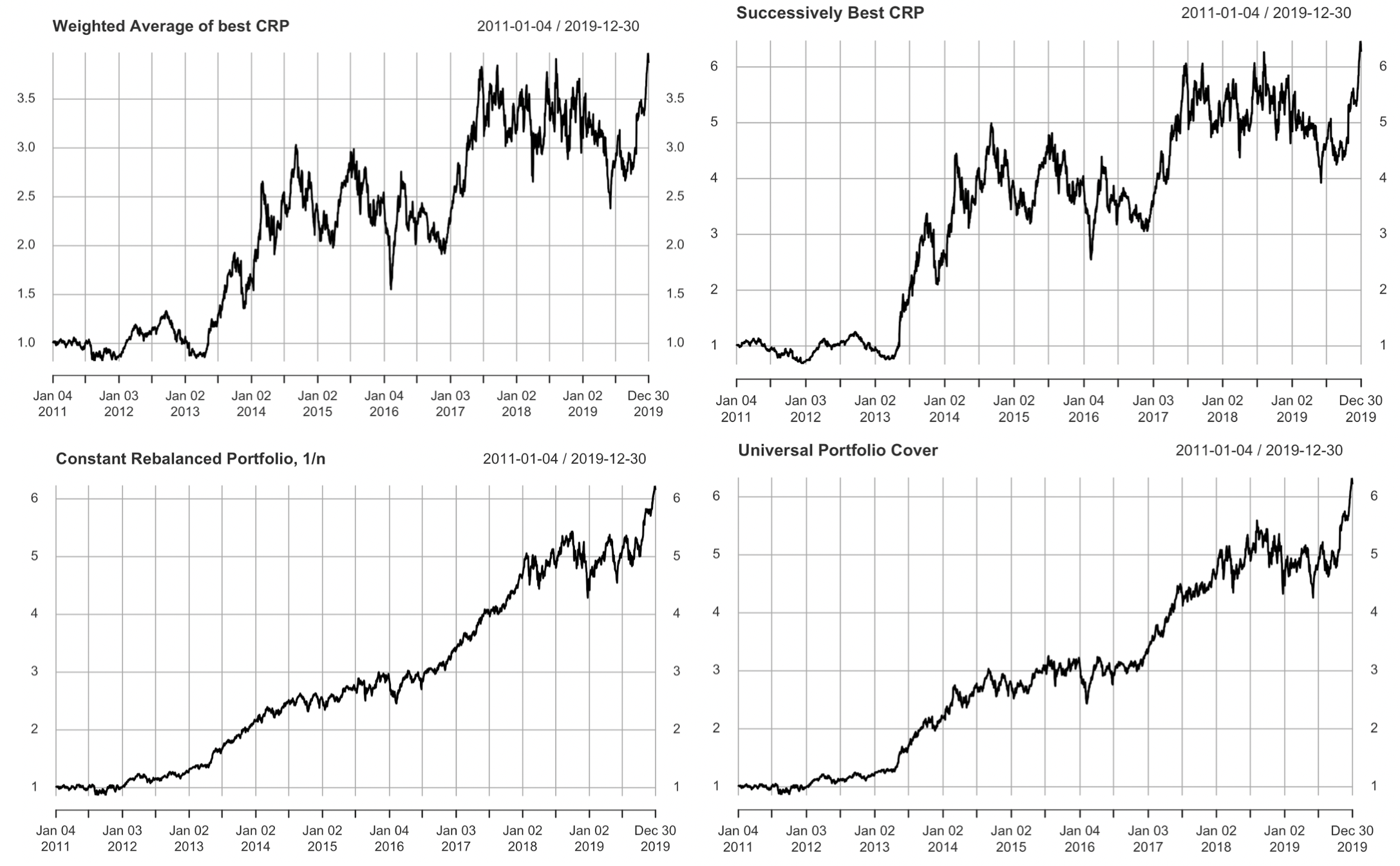}
  \caption{Comparison of CRP based on universal portfolio algorithm} 
\end{figure}

\subsubsection{Four Basic Models}

\textit{Markowitz Modern Portfolio Model} is a portfolio optimization model, emphasizing the inherency of risk. It uses historical return and risk as reference, and helps select the most effective portfolio. Using this model, we can construct an efficient frontier of optimal portfolios offering the maximum possible expected return for a given level of risk.

\textit{Constant Correlation Model} is a mean-variance portfolio selection model, where the correlation of returns between any pair of different securities is considered to be the same. After realizing the past correlation structure hold information about the future average correlation, we predict future correlation with the aggregate technique, by averaging all correlation coefficients in the past correlation structure. Below is the formula we use to calculate correlation matrix:
\begin{equation}
    \rho = \frac{\sum_{i=1}^N\sum_{j=1}^N\rho_{ij}}{\frac{N(N-1)}{2}}
\end{equation}

\textit{Single Index Model} is a regressive model considering the market performance. Our assumption is that all the securities are related to the market index as a whole, so here we set S\&P500 market return as our index. For each security, we estimate parameter $\alpha_i$ and $\beta_i$ to measure their relationship with market index, which can be represented by the following formula:
\begin{equation}
 R_i = \alpha_i+\beta_i*R_M + \epsilon_i
\end{equation}
The equation shows that the stock return influenced by the market $\beta$ and has a specific firm expected value $\alpha$. 

\textit{Multi Index Model} also perform a regression analysis to describe asset returns. The first factor is the \textit{excess return of the market portfolio}, which is the sole factor in CAPM. The second factor small minus big (SMB), measures the difference in returns on a portfolio of small stocks and a portfolio of big stocks. The third factor high minus low (HML), measures the difference in returns on a portfolio of high book-to-market value (BE/ME) stocks and a portfolio of low BE/ME stocks. \textit{Fama French three factors model} assumes that excess return on the $j^{th}$ asset for the $t^{th}$ holding period is linearly correlated with those three risk factors. The return and risk are estimated below:
\begin{equation}
R_{it}-R_{ft} = \alpha_{it} + \beta_{i1}(R_{Mt}-R_{ft}) + \beta_{i2}F_1 +\beta_{i3}F_2
\end{equation}
where $F_1 \sim$ SMB and $F_2 \sim$ HML

\begin{align*}
\sigma_i^2 &= \beta_{i1}^2 \sigma_{R_M-R_f}^2 + \beta_{i2}^2\sigma_{SMB}^2+\beta_{i3}^2\sigma_{HML}^2 \\
\sigma_{ij} &= \beta_{i1}\beta_{j1} \sigma_{R_M-R_f}^2 + \beta_{i2}\beta_{j2}\sigma_{SMB}^2+\beta_{i3}\beta_{j3}\sigma_{HML}^2  \numberthis \label{eqn}
\end{align*}

We can see from Figure 5, all of four models provide us similar results. During the heyday they can reach around \$5 but at last swing down even below \$1. Generally speaking Constant Correlation Model performs the best but still shows unsatisfying result. Multi-factor Model ends up with \$0.41. Investors certainly can choose to sell it before 2019 is coming. 

\begin{figure}[H]
  \centering
  \includegraphics[width=8cm, height=4.2cm]{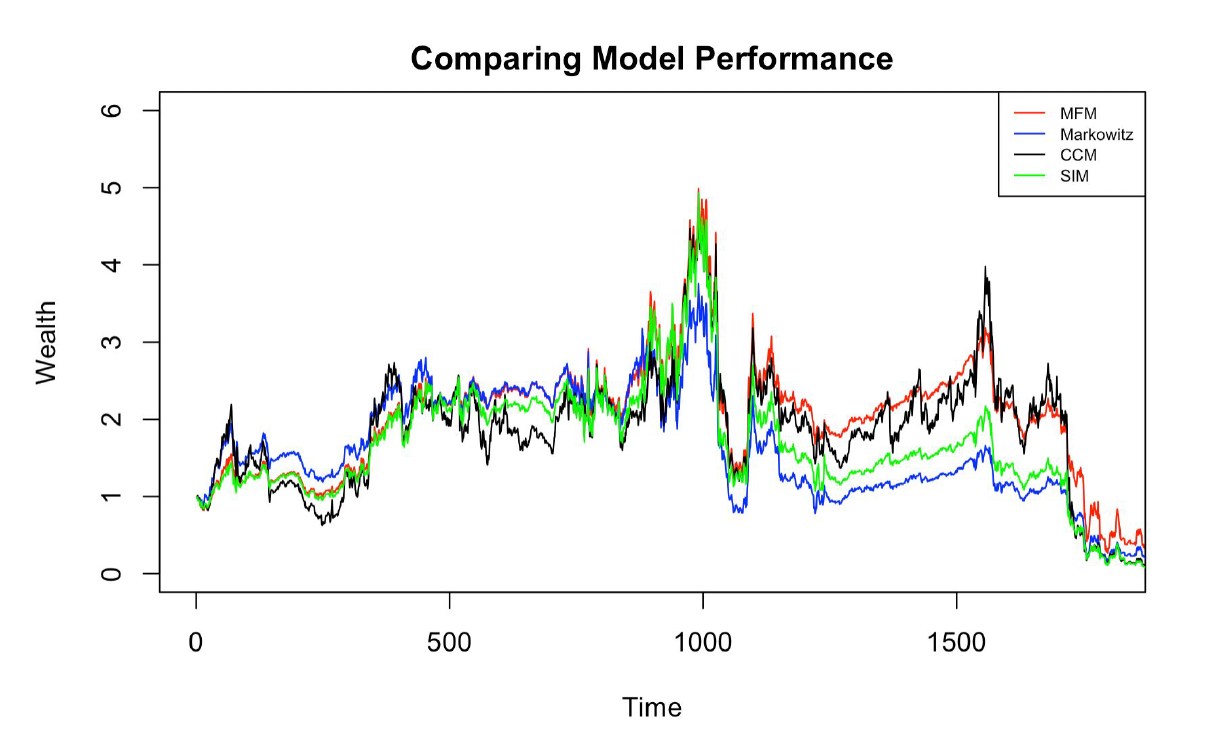}
  \caption{Comparison of Model Performance} 
\end{figure}

\subsubsection{Value at Risk and Expected Shortfall}
As we have already generated four basic models,in order to explore more portfolio risk structure, we utilize two methods, parametric and non-parametric methods. Here, \textit{Value at Risk} (VaR) and \textit{Expected Shortfall} (ES) measure the risk, and \textit{Bootstrap} method confidence interval can capture it more precisely. \par
As for parametric method, we calculate VaR and ES based on assumption that the our stock returns follow t-distribution. By using formulas below, where S is the size of the current position and $\nu$ is the degree of freedom,
\begin{gather*}
   \widehat{VaR}^t(\alpha) = -S \times F_\nu^{-1}(\alpha) \\
   \widehat{ES}^t(\alpha) = \frac{-S}{\alpha} \times \int_{-\infty} ^{F_\nu ^{-1}(\alpha)}xf_\nu(x)dx
\end{gather*} we find out that Multi-factor Model has the lowest VaR and ES which are 0.065 and 0.001 respectively. 
\begin{figure}[H]
  \centering
  \includegraphics[width=8cm, height=1.4cm]{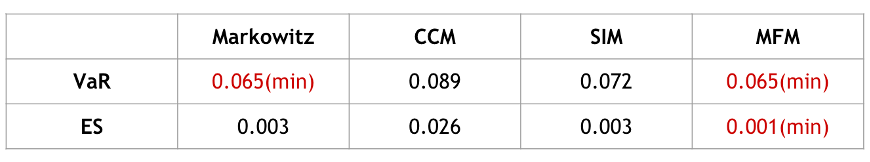}
  \caption{Parametric method result} 
\end{figure}

Non-parametric method is mainly based on the historical performance. By ordering the return and finding the sample quantile, we are able to get the approximate VaR and ES. The formulas are listed below:

\begin{gather*}
   \widehat{VaR}^{np}(\alpha) = -S \times \hat{q}(\alpha) \\
   \widehat{ES}^{np}(\alpha) = -S \times \frac{\sum_{i = 1}^nR_iI\{R_i \leq \hat{q}(\alpha)\}}{\sum_{i = 1}^nI\{R_i \leq \hat{q}(\alpha)\}}
\end{gather*}

It turns out that the Markowitz model has relatively the lowest VaR and ES, which are 0.072 and 0.140 respectively.

\begin{figure}[H]
  \centering
  \includegraphics[width=8cm, height=1.4cm]{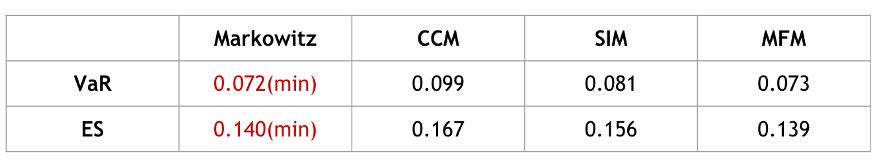}
  \caption{Non-parametric method result} 
\end{figure}

Next, we use PerformanceAnalytics package in R to categorize three Bootstrap methods which are Modified, Gaussian and Historical for confidence interval. 

\begin{figure}[H]
  \centering
  \includegraphics[width=8cm, height=2.4cm]{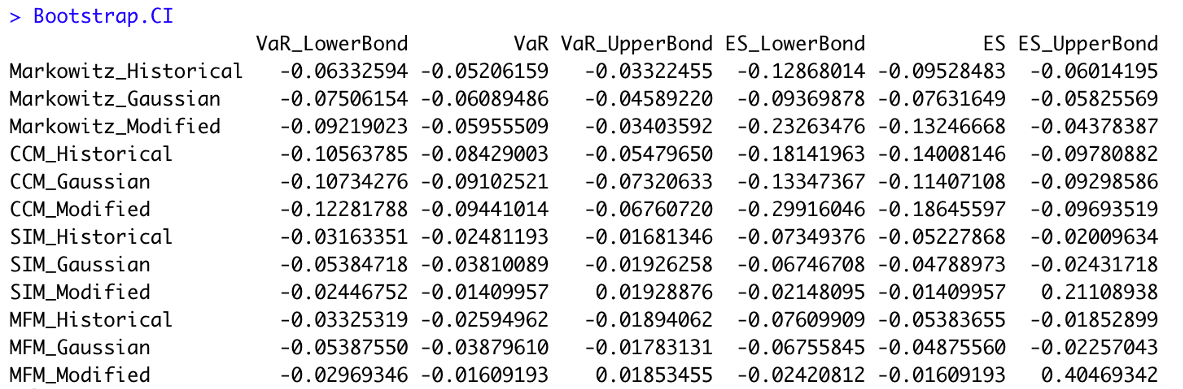}
  \caption{Bootstrap Confidence Interval} 
\end{figure}

\subsection{Return Prediction}
In this section we will go further into return prediction based on factor analysis and time series analysis. Primarily, we split our calculated log return into training and testing set. Training data start from the first trading day of 2012, to the last trading day of 2018, and testing data contain all trading days of 2019. Afterwards, we collect seven factors including Volume, Market Return, Inflation Rate, Risk-free Rate, GDP, CPI and Unemployment Rate, initially. In order to see which factors are significantly contributed to our model, we perform a model selection The result only choose Volume, Market Return and Risk-free Rate as significant factors. \par

One interesting thing about stock return is that the residuals do not satisfy the general assumptions of multivariate regression analysis. Generally we assume that the residuals follow independent identically distributed $\mathcal{N}(0, \sigma^2)$, thus after fitting regression model we are done with predicting process, since the predicted value is just the fitted value due to zero-mean residuals. While this is not the case here. Financial data have the so-called \textit{cluster effect} that observations do not follow linear pattern but rather tend to cluster due to heteroskedasticity. Therefore, the conclusions and predicted value one can draw from the model will not be reliable, which motivate us to use the \textit{GARCH} model to capture the volatility variations. Based on the log-return residuals of the four models and their ACF and PACF plots performance, we have selected different ARMA+GARCH model respectively. \par
Comparing prediction result, Multi-Factor Model has adjusted R-squared 0.2503, so the best we can do is merely to explain 25.03\% of the variation in the calculated log return. 

\end{multicols}

\begin{table}[H]
\centering
 \begin{tabular}{| c | c | c |} 
 \hline
 Models & Factor Model & Residual Model \\ 
 \hline
 MM & $9.28 \times 10^{-3} - 2.73 \times 10^{-12} V + 1.47 R_m - 3.48\times 10^{-4}R_f  + \epsilon$ & ARMA(7,7)+GARCH(1,1)\\ 
 \hline
 CCM & $1.37 \times 10^{-2} - 4.01 \times 10^{-12} V + 1.89 R_m - 5.68\times 10^{-4}R_f  + \epsilon$ & ARMA(3,3)+GARCH(1,1)\\ 
 \hline
 SIM & $1.01 \times 10^{-2} - 3.01 \times 10^{-12} V + 1.56 R_m - 3.56\times 10^{-5}R_f  + \epsilon$ & ARMA(3,3)+GARCH(1,1)\\ 
 \hline
 MFM & $7.92 \times 10^{-3} - 2.35 \times 10^{-12} V + 1.51 R_m - 4.84\times 10^{-5}R_f  + \epsilon$ & ARMA(3,3)+GARCH(1,1)\\ 
 \hline
\end{tabular}
\caption{Factor Models and Residual Models for four series of log-returns}
\label{table:1}
\end{table}

\begin{multicols}{2}

\begin{figure}[H]
  \centering
  \includegraphics[width=8cm, height=5cm]{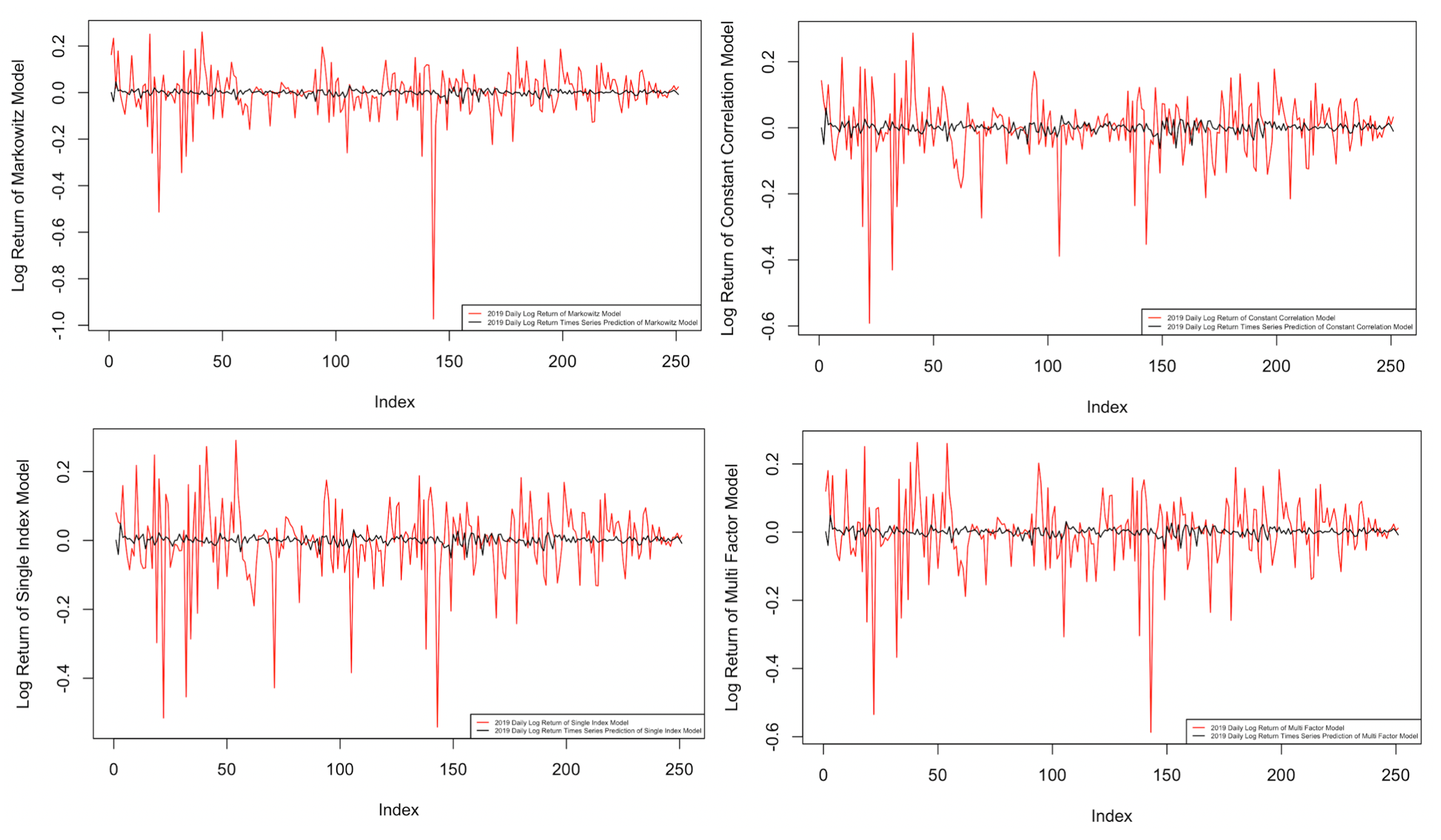}
  \caption{Actual return (in red) and fitted return (in black) from four models} 
\end{figure}
Although the outcomes of return prediction are not extraordinary, in fact, it is par for the course. To conduct further analysis, there are a few potential reason lie behind.
\begin{itemize}
    \item Portfolio log return based on four models calculated is not an actual asset but a virtual one. It may not fit perfectly based on real world data analysis
    \item Using estimated mean of log return as our prediction may not be appropriate since possible oscillation exists
    \item There are also other hidden factors that we did not captured and what would actually have influences on the stock returns are inscrutable
    
\end{itemize}

\section{Conclusion}
Return and risk trade-off is an eternal theme, that investors always think about. In this article are are focusing on dealing with the relationship between them, by utilizing most famous Markowitz's Modern Portfolio Theory. In addition to that, under four models, Markowitz Model, Constant Correlation Model, Single Index Model and Multi-factor Model, we develop efficient investment strategy, construct portfolios, thereby seize the volatility structure and predict future performance. All in all, Multi-factor Model seems to be the outstanding one in both risk management and return prediction, however, the final result is still embarrassing. \par
As you may wonder, why cannot the models capture the reality happen in the real world well? We believe that financial market has more variability beyond the scope of the whole bunch of fundamental theory. After further study by analyzing more complex model we could do better.

\section{Further Improvement}
If time permitted, we are planning to finish model construction based on Stochastic Portfolio Theory, with well-defined machine learning tools mentioned by Vervuurt and Kom Samo.  \par
We also made attempt on using copula to fit multivariate joint distribution, based on \textit{Sklar's Theorem}, which states that a collection of marginal distributions can be coupled together via a copula to form a multivariate distribution. A copula is a multivariate CDF whose univariate marginal distributions are all Uniform(0,1).\cite{copula} Based on the goodness of fit test we choose to use \textit{t-copula} and generate the final multivariate distribution in dimension 8 by t-copula with degree of freedom 11. The result is unsatisfying, determined by \textit{backtesting}. In the later study we could investigate what happened in copula utilization. \par

\end{multicols}

\clearpage
\appendix
\section{Appendix}

\begin{algorithm}[H]
\SetAlgoLined
\KwResult{Based on CAPM, calculate weight and corresponding wealth for each trading days}
 i. Initialize total wealth = \$1, each stock's weight = $\frac{1}{8}$ \;
 ii. Use $R_f$, $R_i$ and $\sigma_i$ based on four models, calculate the tangent portfolio which maximize the Sharpe's Ratio as optimal one, derive the weights and final earnings at the end of trading day, and reinvest it at the beginning of next trading day with calculated weight\;
 iii. Iterate (ii) until the end of the trading period, we get a weight matrix \textbf{W} with each row representing daily weights and a wealth vector \textbf{S} storing the daily earnings \;
 iv. Return the last entry in \textbf{S} as our final result, and plot evolution of wealth to visualize
 \caption{Evaluate Basic Model Performance}
\end{algorithm}

\begin{algorithm}[H]
\SetAlgoLined
 i.	Simulate 500 bootstrap samples consisting of log-returns based on our four models\;
 ii. Calculate bootstrap sample $\widehat{VaR}$ and $\widehat{ES}$ based on simply just quantile (Historical Method)\;
 iii. Assume log-return approximately normal and compute $\widehat{VaR}$ and $\widehat{ES}$ (Gaussian Method)\;
 iv. Calculate $\widehat{VaR}$ and $\widehat{ES}$ by using Cornish-Fisher Expansion (Modified Method)\;
 v. Order the 500 $\widehat{VaR}$ and $\widehat{ES}$, calculate the sample quantile $\hat{q}_{0.025}$ and $\hat{q}_{0.975}$, respectively\;
 vi. Lower = $2\widehat{VaR}(\widehat{ES}) - \hat{q}_{0.975}$, upper = $2\widehat{VaR}(\widehat{ES}) - \hat{q}_{0.025}$
 
 \caption{Bootstrap Confidence Interval Construction}
\end{algorithm}

\begin{multicols}{2}

\begin{figure}[H]
  \centering
  \includegraphics[width=8cm, height=5cm]{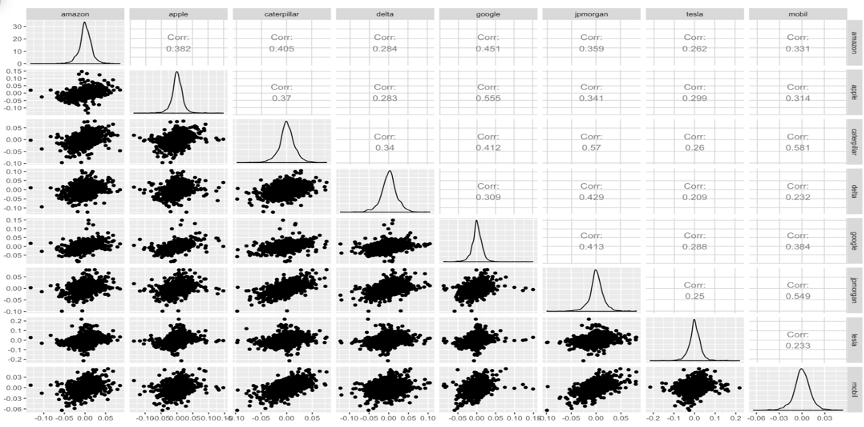}
  \caption{Correlation plot for 8 stocks} 
\end{figure}

\begin{figure}[H]
  \centering
  \includegraphics[width=5cm, height=5cm]{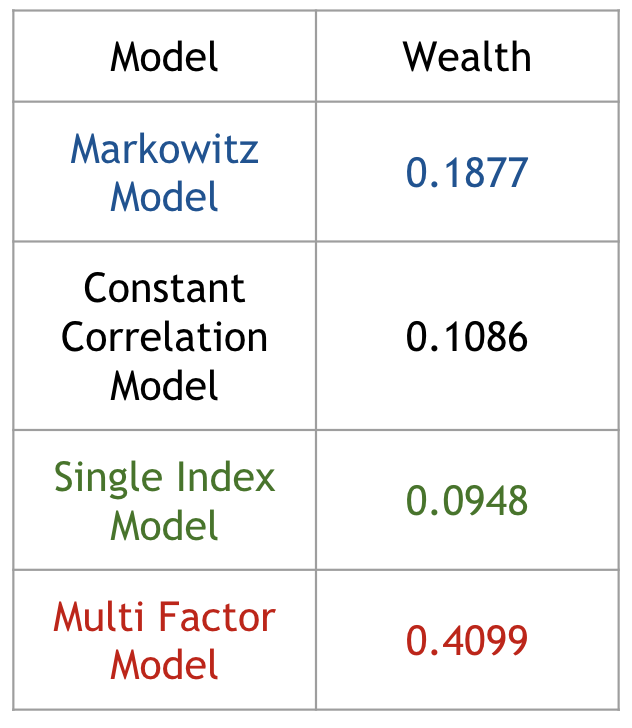}
  \caption{Final performance for four models} 
\end{figure}

\begin{figure}[H]
  \centering
  \includegraphics[width=8cm, height=5cm]{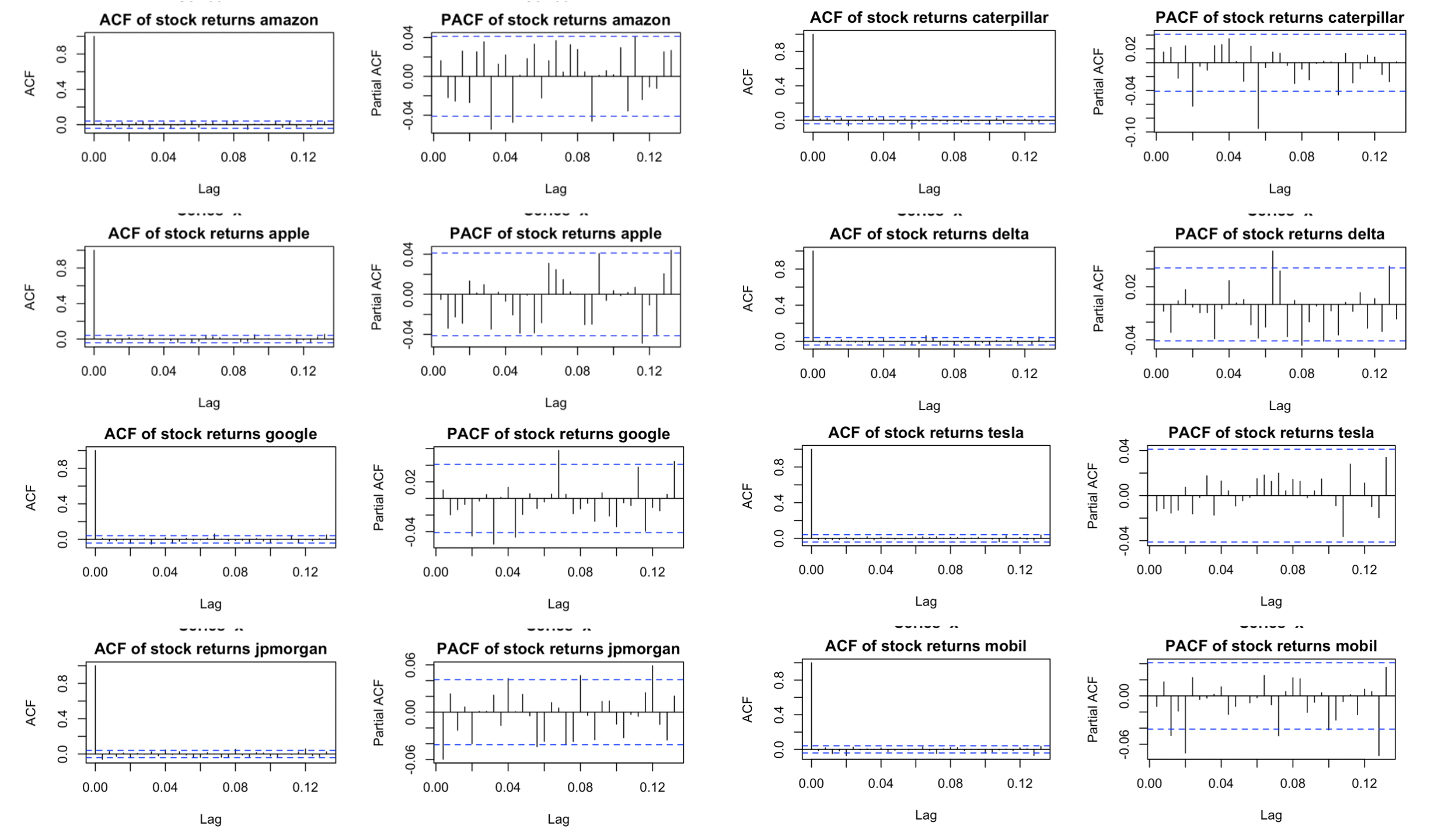}
  \caption{Sample ACF and PACF plot} 
\end{figure}

\begin{figure}[H]
  \centering
  \includegraphics[width=8cm, height=5cm]{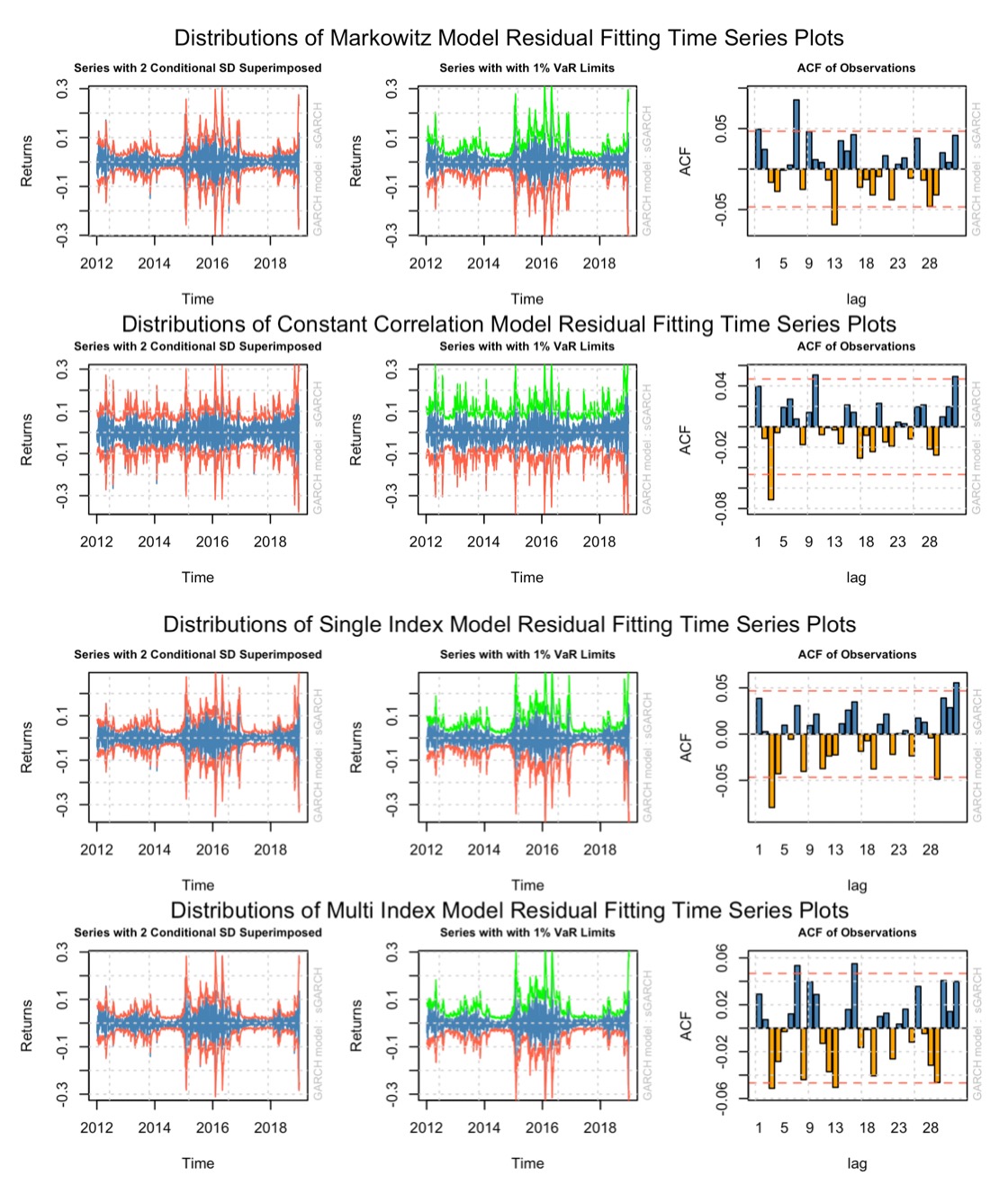}
  \caption{Time series evaluation} 
\end{figure}

\end{multicols}

\end{document}